\documentclass{elsart}

\usepackage{graphicx}

\usepackage{amssymb}

\begin{document}

\begin{frontmatter}



\title{Cosmology from Planck}


\author{Amedeo Balbi}

\address{Dipartimento di Fisica, Universit\`a di Roma ``Tor Vergata''\\
and INFN, Sezione di Roma ``Tor Vergata''\\
Via della Ricerca Scientifica 1, 00133 Roma, Italy}

\begin{abstract}
I briefly review some of the main scientific outputs expected from the upcoming Planck mission. Planck will map the CMB sky with 5' resolution and $\mu$K sensitivity, with minimal foreground contribution and superb control on systematics. It will collect the entire information enclosed in the temperature primary anisotropy signal and will also get a good measurement of the polarized component of the CMB.  This will have profound implications on our knowledge of the physics of the early universe and on the determination of cosmological parameters.
\end{abstract}


\end{frontmatter}

\section{Introduction: the CMB circa 2006}

\begin{figure}
\centering
\includegraphics[width=\columnwidth]{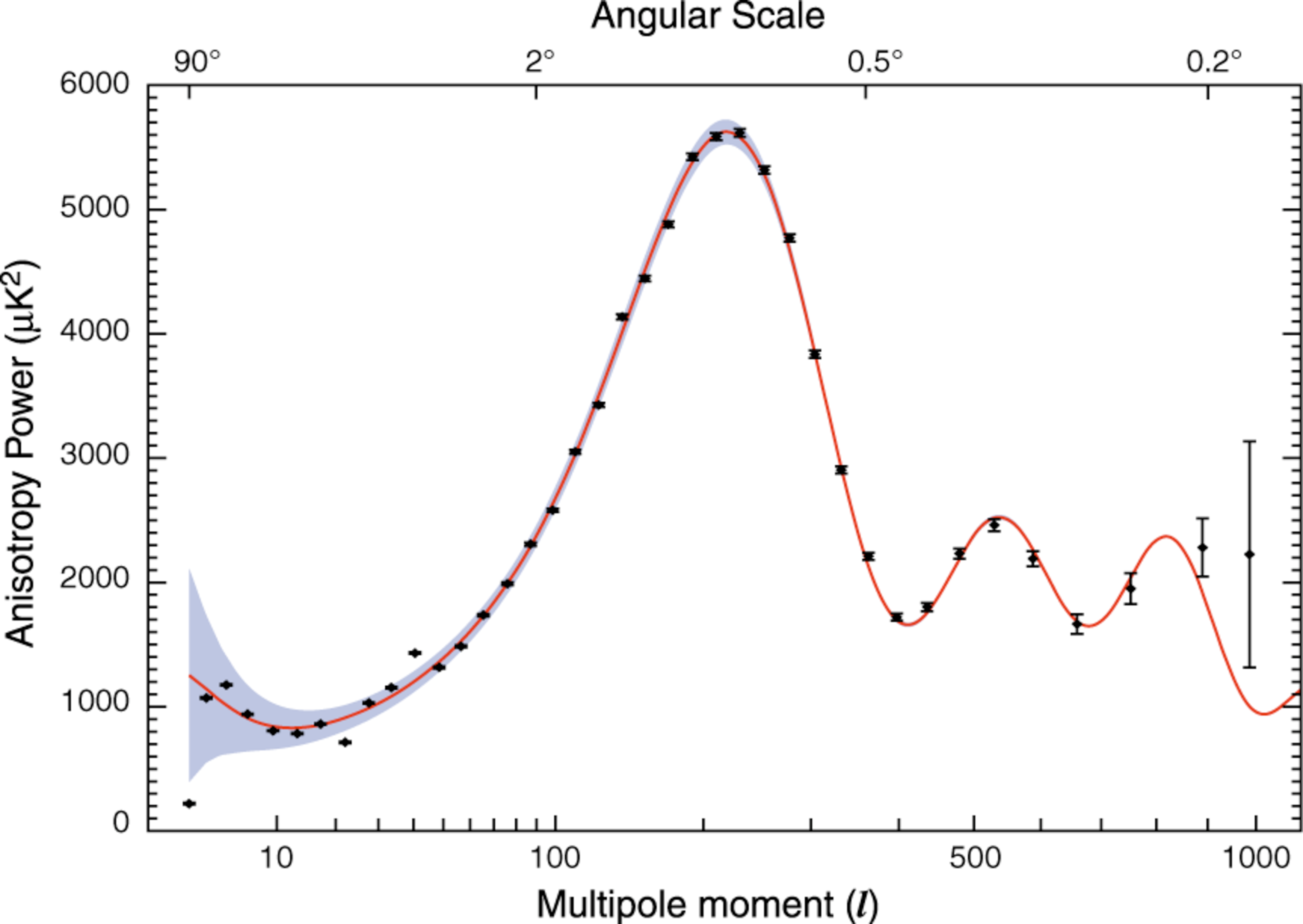}
\caption{The CMB temperature power spectrum observed by WMAP (from \cite{wmap}). The continuos line is the $\Lambda$CDM best fit model, and the shaded area is the associated cosmic variance. \label{fig:wmap}}
\end{figure}

The current situation of CMB anisotropy measurements is summarized in Figure~\ref{fig:wmap}. We have an exquisitely accurate reconstruction of the temperature anisotropy angular power spectrum from the WMAP satellite \cite{wmap}, which is essentialy cosmic variance limited up to the first acoustic peak. The second peak is also determined quite well. What this tells us is that acoustic oscillations actually took place in the primeval plasma and that inflationary adiabatic perturbations are the best candidate to explain structure formation in the universe by gravitational instability. We also have a very good measurement of the baryon content ($\Omega_b h^2=0.0223^{+0.0007}_{-0.0009}$ \cite{Spergel et al. 2006}) from the ratio of first to second peak heights, in agreement with the primordial nucleosynthesis predictions of the observed light nuclei abundances. The acoustic angular scale at recombination is determined with high precision ($\theta_A=0.595\pm 0.002$ degrees \cite{Spergel et al. 2006}) and is providing us with strong evidence that we live in a nearly flat universe. Overall, a flat model with only six free parameters (the densities of baryons and cold dark matter, the amplitude and spectral index of scalar density perturbations, the present value of the Hubble parameter, and the optical depth to reionization) is a very good fit to the current CMB data.

Polarization information has also started to be collected, although we are still very far from the accuracy attained by temperature data. The B2K experiment \cite{montroy_b2k} recently measured the EE and TE power spectra, and the same did WMAP \cite{wmap_yr3_pol}. What these first results are telling us is that there is a consistency between the predicted level of polarized signal in the standard cosmological model and what we actually observe. This reinforces the case for adiabatic perturbations and inflation. Polarization measurements, however, are still too uncertain to provide constraints on cosmological parameters with an accuracy comparable to those coming from temperature data.

When we look at these wonderful results, we are confronted with the unavoidable question: what remains to be done after WMAP? The answer is: quite a lot, actually. 

\section{What after WMAP?}

To put things in the right context, we can quickly go through a list of some of the major questions left fully or partially unanswered by WMAP. First, there is a need for improved tests of inflation, through better constraints on the spectral index of scalar perturbations and the measurement of its variation (or lack thereof) with $k$ (i.e.\ running). WMAP seemed to show hints of deviations from a Harrison-Zel'dovich spectrum ($n_s=1$) although this is still within the 2$\sigma$ bound \cite{Kinney et al. 2006}. A significant departure from simple scale invariance would be a smoking gun for realistic inflationary models. 

Accurate measurements of matter density from the CMB are still lacking. This is due to the fact that the information on the breakdown of total matter content into baryons and dark matter can only be done when the third acoustic peak is reconstructed with an accuracy comparable to that of the first two. This is definetely not possible with current data. The uncertainty on $\Omega_m h^2$ affects the determination of other parameters, including curvature (since this can be measured from the acoustic angular scale only if we also know the distance to last scattering, which depends on $\Omega_m$ and $h$).

Then we enter into polarization muddy waters. We would very much need a measurement of the EE power spectrum, since this would help with the determination of the detailed reionization history. This, in turn, would lead to improved constraints on the scalar spectral index, by removing its degeneracy with the optical depth. Measuring B mode would also be nice, since this would tell us about a tensor mode contribution to primordial perturbations, leading to a better discrimination of inflationary model. 

Finally, there is interesting science hidden in the contribution from secondary anisotropy (for example from the Sunyaev-Zel'dovich effect in clusters of galaxies, weak lensing of the large scale structure on the CMB photons, and so on), which is expected to peak at much smaller scales than those accessible to WMAP. There is also the issue of testing primordial non Gaussianity (another important signature of specific models of inflation) through higher order statistics in multipole space (such as the bispectrum, trispectrum and so on) or analysis in the pixel domain (such as those based on Minkowski functionals): the latter would greatly benefit from high signal-to-noise maps of the CMB. 

These goals will definitely be within the capabilities of the upcoming Planck mission.

\section{The Planck satellite}

The Planck ESA's satellite\footnote{http://www.rssd.esa.int/Planck/} is currently scheduled for launch in August 2007. It will be the first mission to map the entire sky at a resolution as high as $5'$ with $\mu$K sensitivity (three times better resolution than WMAP, and an order of magnitude lower noise at 100 GHz). The frequency coverage will also be impressive, spanning the range $30-857$ GHz with 9 channels and two different detector techniques (HEMT arrays for the Low Frequency Instrument and bolometers arrays for the High Frequency Instrument). This will allow for optimal foreground separation and for superb control on systematics. The technical details of the Planck mission are described elsewhere: see for example the Planck Blue Book \cite{bluebook}, which also illustrates the science goals that Planck is expected to achieve. In the following I will summarize what seem to me the most important assets of Planck, particullarly when compared to the recent WMAP results.

\subsection{Science from Planck temperature measurements}

\begin{figure}
\centering
\includegraphics[width=\columnwidth]{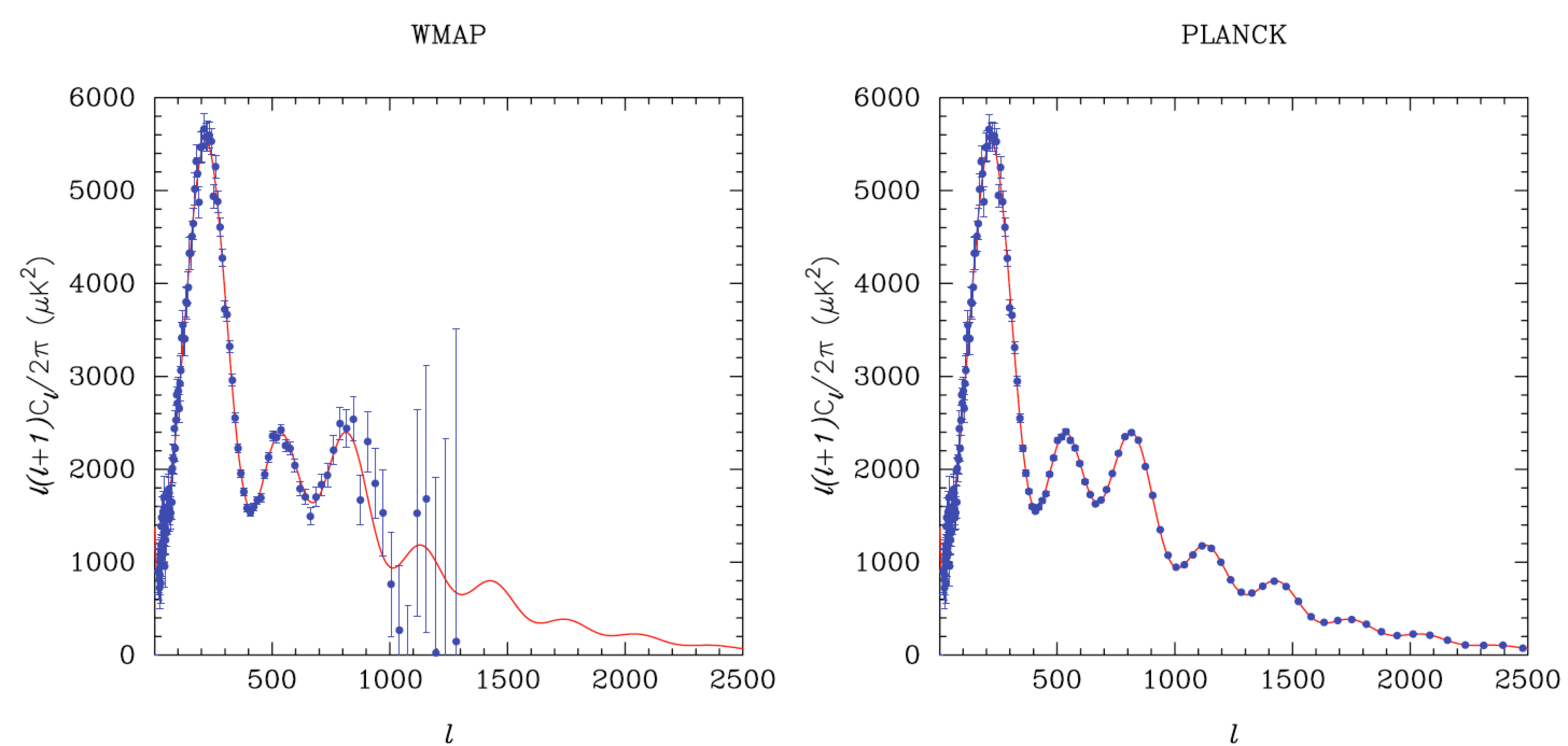}
\caption{Comparison between the CMB power spectrum measurements expected from WMAP 4 year data (left panel) and from Planck (right panel). The underlying cosmological model is the $\Lambda$CDM WMAP best fit. (From \cite{bluebook}).  \label{fig:planck}}
\end{figure}

Planck will essentially be the definitive mission as far as primary CMB temperature anisotropy is concerned. The reconstructed power spectrum will be cosmic variance limited up to the fourth acoustic peak, and will get impressive accuracy at even higher multipoles, well into the anisotropy damping tail. This will have important consequences on cosmology. A forecast of how the temperature power spectrum of the $\Lambda$CDM concordance model will be reconstructed by Planck is shown in Figure~\ref{fig:planck}.

\begin{figure}
\centering
\includegraphics[width=\columnwidth]{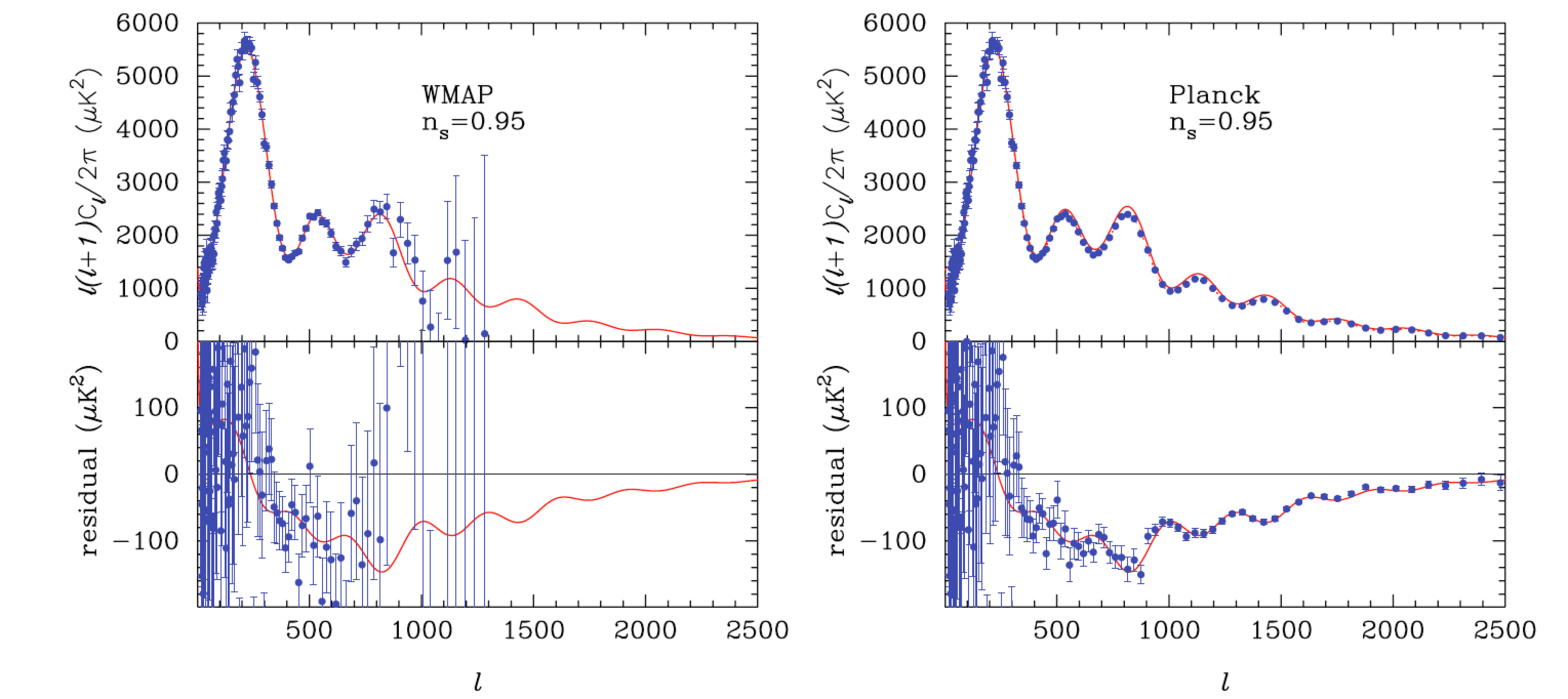}
\includegraphics[width=0.98\columnwidth]{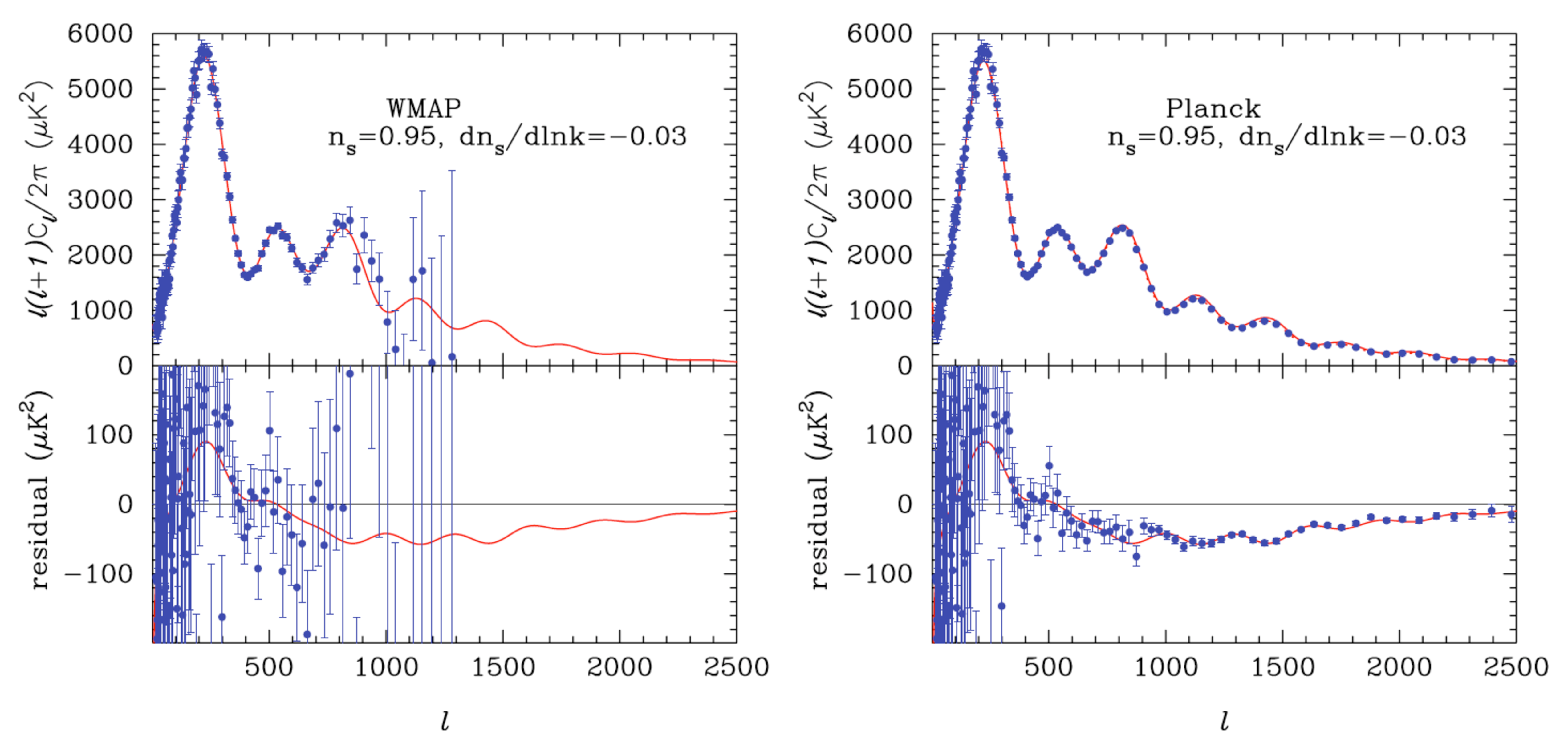}
\caption{Reconstruction of the spectral index of primordial scalar perturbations. In the upper panels, the continuous line is a $\Lambda$CDM concordance model with scale invariant spectrum ($n_s=1$). The points are simulated observations from WMAP (left panel) and Planck (right panel) of the same cosmological model, except that $n_s=0.95$. The residuals (showed below the two plots) quantify the ability of the two experiments to discriminate different values of the spectral index. In the lower panels, the same exercise is performed assuming no running of the spectral index (continuous line) and a run of $dn_s/d\ln k=-0.03$ (simulated data points). (From \cite{bluebook}). \label{ns}}
\end{figure}

Figure~\ref{ns} shows just one example of the giant step forward that the Planck data will represent. The spectral index of scalar primordial perturbations and its running will be measured by Planck with great precision, allowing to distinguish among different models of inflation. 

The longer lever arm in multipole space will also be crucial to determine the quantity of matter in the universe. The parameter $\Omega_m h^2$, which affects the third peak height relative to the first and second, will be measured to better than $1\%$ accuracy \cite{bluebook} (compared to the current $7\%$ \cite{Spergel et al. 2006}). This will allow increased accuracy on the determination of the distance to last scattering. Combining this with the precise measurement of the angular diameter distance to recombination, one will firmly establish the geometrical properties of the universe, with less need for prior assumptions.

Although CMB temperature anisotropy is not the primary tool to investigate the properties of the mysterious dark energy component which makes up the $70\%$ of the universe (see, e.g.\ \cite{Riess et al. 2004}), it can be a valuable complementary tool, in at least two ways. First, by accurately determining the curvature of the universe it can remove geometrical degeneracies and serve as a strong prior for other astrophysical observations (see, e.g.\ \cite{Balbi et al. 2003}). Second, it can be correlated to tracers of the large scale structure (such as galaxy redshift surveys) to detect variations of the gravitational potential occurring during the transition from matter to dark energy domination, through the integrated Sachs-Wolfe effect \cite{Sachs & Wolfe 1967} (for an example of current results obtained with this tool see, e.g.\ \cite{Pietrobon 2006} and references therein). The Planck maps, with their high signal-to-noise and minimal foreground contamination, will be an excellent target for this kind of analysis.

\subsection{Science from Planck polarization measurements}

\begin{figure}
\centering
\includegraphics[width=\columnwidth]{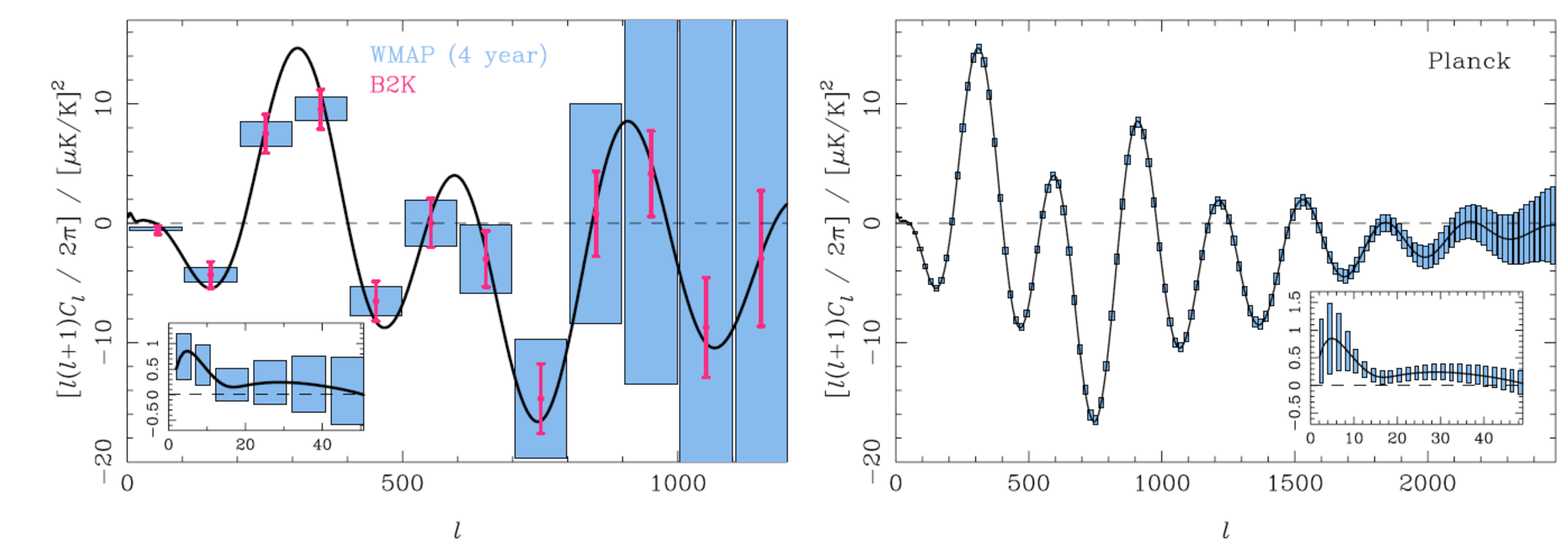}
\includegraphics[width=\columnwidth]{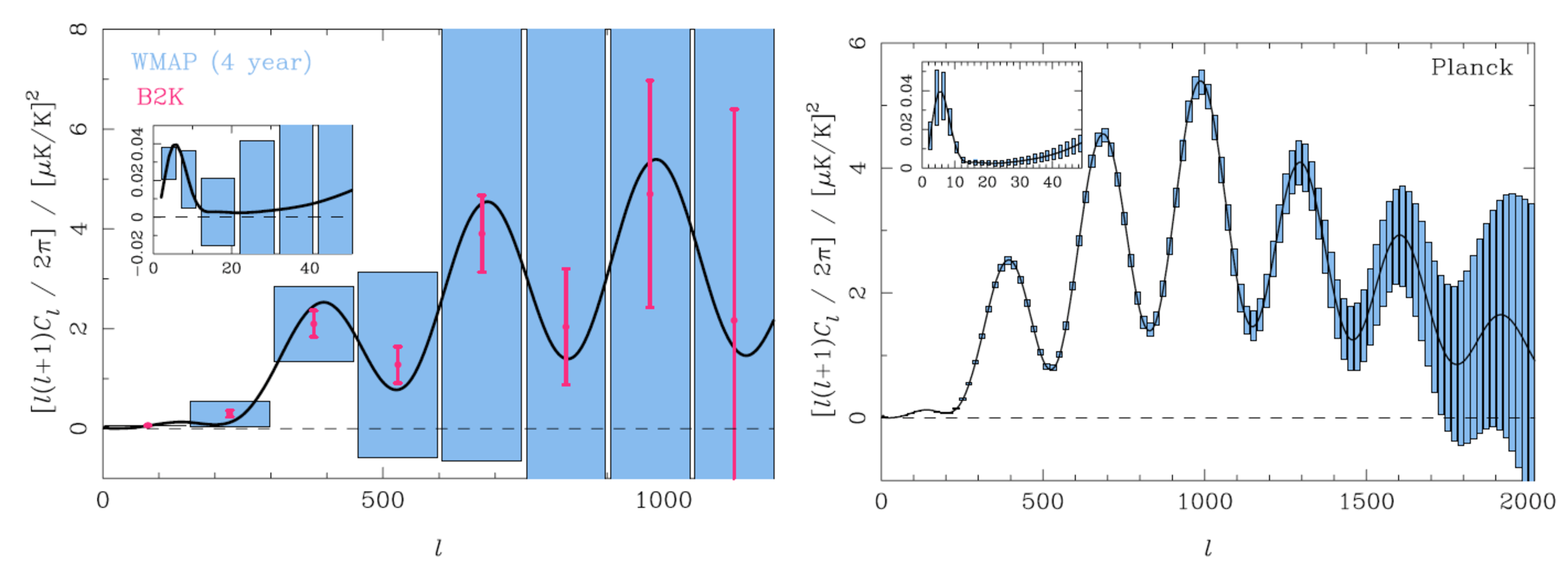}
\caption{Expected measurements of the CMB TE (upper panels) and EE (lower panels) power spectrum from WMAP and B2K (left panels) and Planck (right panels). The underlying theoretical model is a $\Lambda$CDM with  tensor-to-scalar ratio $r=0.1$ and optical depth to reionization $\tau=0.17$. (From \cite{bluebook}). \label{planckpol}}
\end{figure}

Although not specifically designed for polarization measurements, Planck will nevertheless be able to determine the EE, BB and TE angular power spectra with good accuracy. Figure~\ref{planckpol} shows the forecasted TE and EE power spectrum for Planck. 

The are several major breakthroughs than can be expected from accurate measurements of the polarized component of the CMB anisotropy. 
The predicted polarization EE power spectrum shows the same kind
of acoustic features that are now well known to exist in the
temperature anisotropy spectrum. The peaks in polarization, however,
are out of phase with those in the temperature: the EE spectrum
has maximum intensity were the temperature is at a minimum, and
viceversa. This is due to the fact that polarization is generated by
quadrupolar anisotropy at last scattering, and this is closely related
to the velocity of the coupled photon-barion fluid. The maximum
compression or rarefaction (and minimum velocity) of the fluid
corresponds to peaks in the temperature anisotropy (and troughs in the
polarization). For the same reason, the TE cross-correlation power
spectrum shows pronounced peaks corresponding to the interleaved sets of maxima and minima in the two separate components. A detection of such features in the polarization and cross-polarization power spectra would then be crucial: it would give an independent confirmation of the fact that acoustic oscillations took place in the primeval plasma. This would strengthen the case for the adiabatic nature of primordial perturbations thus providing strong support to cosmic inflation.

\begin{figure}
\centering
\includegraphics[width=\columnwidth]{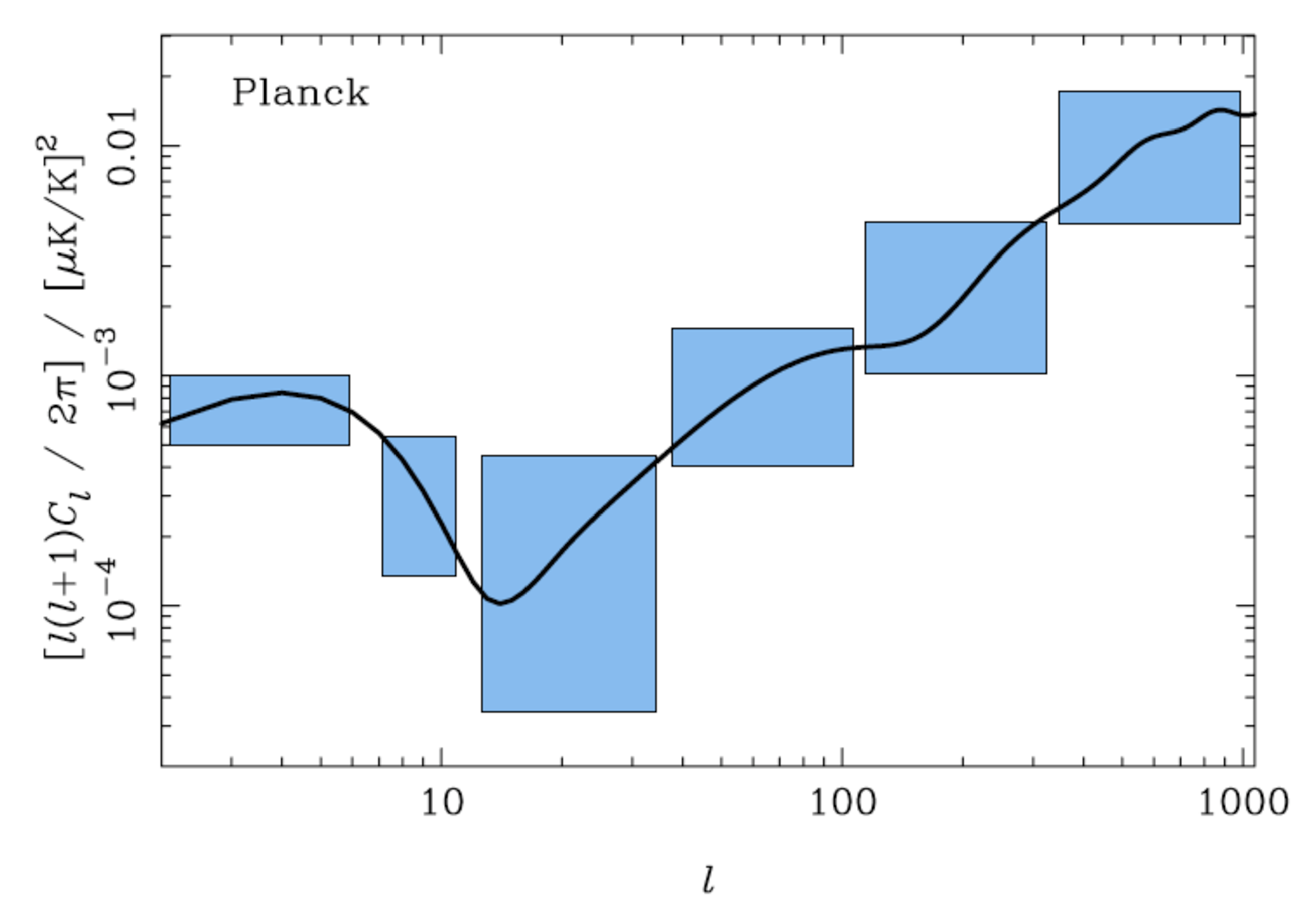}
\caption{Expected measurements of the BB power spectrum from Planck, assuming the same underlying model of Figure~\ref{planckpol} \label{bmode} (from \cite{bluebook}).}
\end{figure}

Since it probes the epoch of decoupling, polarization allows one to perform detailed tests of the recombination physics. In particular, it is a well known fact that the universe underwent a phase of reionization at redshifts of at least
$z\sim 5$, during the early stages of structure formation. The investigation of these so-called dark ages is an active subject of investigation \cite{Choudhury & Ferrara 2006}. When photons are diffused by free electrons along the line of sight, the CMB temperature anisotropy signal gets damped. The amount of damping would be a powerful probe of the optical depth to reionization. However this effect is masked by other physical mechanisms. For example, there is a strong degeneracy between the optical depth and the spectral index of primordial perturbations. CMB polarization would prove very powerful in breaking this degeneracy: if the optical depth is non zero, a recognizable polarization signature gets generated at large angular scales,
allowing investigation of the detailed reionization history, e.g.\ by discriminating models that have the same optical depth but a different evolution of the ionization fraction with redshift \cite{Hu & Holder 2003}. Not only the characterization of the detailed ionization history of the universe would have a strong scientific impact in itself, but it would also greatly increase the accuracy of the determination of other cosmological parameters, such as the above mentioned spectral index of scalar primordial perturbations. This, in turn, would be extremely important when constraining models of inflation.

One crucial aspect of CMB polarization has to do with the properties
of its B component. A forecast of the BB power spectrum as should be observed by Planck is shown in Figure~\ref{bmode}. The consequences of detecting B polarization for theoretical models would be enormous. B modes can only be generated when a tensor component of primordial perturbations is present (namely, a background of primordial gravitational waves). This is precisely one of the predictions of inflationary models. The ratio of scalar to tensor fluctuations $r$, is a direct signature of the inflation energy scale, and is strongly related to the amplitude of B modes. Knowledge of $r$ and of the spectral index of the perturbation scalar component can be translated into a measurement of the inflationary slow-roll parameters, thus allowing to discriminate specific models of inflation.

\section{Conclusions}

In the next few years, we can expect a huge breakthrough in our knowledge of cosmology from the Planck mission results. We will be able to extract the complete information embedded in the temperature primary anisotropy data and have good (although not optimal) reconstruction of the polarized component of the CMB anisotropy.  Foregrounds contamination will be minimal and systematics will be under control. This will allow percent accuracy on the measurements of cosmological parameters and will be used as an invaluable source of prior knowledge for upcoming probes of the local universe. 



\end{document}